\documentclass[12pt]{article}

\usepackage[superscript]{cite}
\usepackage[labelsep=space]{caption}

\usepackage{times}

\newenvironment{sciabstract}{%
\begin{quote} \bf}
{\end{quote}}

\newcounter{lastnote}

\title{The Ginger-shaped Asteroid 4179 Toutatis: New Observations from a Successful Flyby of Chang'e-2}

\author
{Jiangchuan Huang$^{1,3*}$, Jianghui  Ji$^{2*}$, Peijian Ye$^{1}$, Xiaolei Wang$^{4}$,\\
Jun Yan$^{5}$, Linzhi Meng$^{3}$, Su  Wang$^{2}$, Chunlai Li$^{5}$, \\
Yuan  Li$^{6}$, Dong Qiao$^{7}$, Wei Zhao$^{8}$, Yuhui   Zhao$^{2}$, Tingxin Zhang$^{1}$,\\
Peng Liu$^{8}$, Yun  Jiang$^{2}$, Wei Rao$^{3}$, Sheng  Li$^{9}$, \\
Changning Huang$^{10}$, Wing-Huen Ip$^{6,11}$, Shoucun  Hu$^{2}$, Menghua  Zhu$^{6}$, \\
Liangliang Yu$^{2}$, Yongliao Zou$^{5}$, Xianglong Tang$^{8}$, Jianyang  Li$^{12}$,\\
Haibin Zhao$^{2}$, Hao Huang$^{3}$, Xiaojun Jiang$^{5}$, \& Jinming Bai$^{13}$\\
\footnotesize{$^{1}$China Academy of Space Technology, Beijing 100094, China}.\\
\footnotesize{$^{2}$Key Laboratory of Planetary Sciences, Purple Mountain Observatory, Chinese Academy of Sciences, Nanjing 210008, China}.\\
\footnotesize{$^{3}$Institute of Space System Engineering, Beijing 100094, China}.\\
\footnotesize{$^{4}$Beijing Institute of Control Engineering, Beijing 100190, China}. \\
\footnotesize{$^{5}$National Astronomical Observatories, Chinese Academy of Sciences, Beijing 100012, China}.\\
\footnotesize{$^{6}$Space Science Institute, Macau University of Science and Technology, Taipa, Macau}. \\
\footnotesize{$^{7}$Beijing Institute of Technology, Beijing 100081,   China}. \\
\footnotesize{$^{8}$Harbin  Institute of Technology,  Harbin  150001, China}.  \\
\footnotesize{$^{9}$School of Electronic Engineering and Computer Science, Peking University, Beijing 100871, China}.\\
\footnotesize{$^{10}$Beijing Institute Of Space Mechanics and Electricity, Beijing 100076, China}. \\
\footnotesize{$^{11}$Institute of Astronomy, National Central University, Taoyuan, Taiwan}. \\
\footnotesize{$^{12}$Planetary Science Institute, AZ 85719, USA}. \\
\footnotesize{$^{13}$Yunnan Astronomical Observatory, Chinese Academy of Sciences, Kunming 650011, China.} \\
\footnotesize{$^{*}$~The authors contribute equally to this work. To
whom correspondence should be addressed. E-mail:jijh@pmo.ac.cn} }

\date{}

\begin{document}


\baselineskip24pt


\maketitle
\clearpage

\begin{sciabstract}
On 13 December 2012, Chang'e-2 conducted a successful flyby of the
near-Earth asteroid 4179 Toutatis at a closest distance of 770 $\pm$
120 meters from the asteroid's surface. The highest-resolution
image, with a resolution of better than 3 meters, reveals new
discoveries on the asteroid, e.g., a giant basin at the big end, a
sharply perpendicular silhouette near the neck region, and direct
evidence of boulders and regolith, which suggests that Toutatis may
bear a rubble-pile structure. Toutatis' maximum physical length and
width are (4.75 $\times$ 1.95 km) $\pm$10$\%$, respectively, and the
direction of the +$z$ axis is estimated to be (250$\pm$5$^\circ$,
63$\pm$5$^\circ$) with respect to the J2000 ecliptic coordinate
system. The bifurcated configuration is indicative of a
contact binary origin for Toutatis, which is composed of two lobes
(head and body). Chang'e-2 observations have significantly improved
our understanding of the characteristics, formation, and evolution
of asteroids in general.

\end{sciabstract}

\section*{Introduction}

Chang'e-2, the second Chinese probe dedicated to the exploration of
the Moon, was launched on 1 October 2010 at 10:59:57 UTC. After it
accomplished its primary objective, an extended mission was designed
for Chang'e-2 to travel to the Sun-Earth Lagrangian point (L2) and
then to an asteroid. On 9 June 2011, Chang'e-2 departed its lunar
orbit and headed directly to L2, where it arrived on 25 August 2011
and began to perform space-environment exploration. The asteroid's
subsequent flyby mission remained under consideration, and
eventually, asteroid 4179 Toutatis was selected as the primary
target for the flyby after considering the leftover fuel of the
spacecraft, the capability of the tracking and control network, and
the fact that the asteroid's closest approach to Earth would occur
on 12 December 2012. An undisclosed story is that Chang'e-2 was
intentionally controlled to move near L2 for more than 230 days to
favor the Toutatis-flyby injection maneuvers. On 15 April 2012,
Chang'e-2 left L2. On 1 June 2012, the spacecraft began its mission
to Toutatis. Chang'e-2 implemented the flyby on 13 December 2012 at
8:29:58.7 UTC at a closest distance of $770\pm 120$ ($3 \sigma$)
meters from Toutatis' surface at a high relative velocity of 10.73
km s$^{-1}$, with highest-resolution images of better than 3 meters
per pixel \cite{Huang13} . As of 14 July 2013, Chang'e-2 is still
traveling into outer space over 50 million kilometers away from
Earth, and it is the first successful multiple-type-objective probe
in history that has ever visited the Moon, a Lagrangian point, and
an asteroid.

Asteroids are remnant building blocks of the formation of the Solar
System. They provide key clues to understanding the process of
planet formation, the environment of the early Solar nebula, and
even the occurrence of life on Earth. Toutatis, as an Apollo
near-Earth asteroid in an eccentric orbit that originates from the
main belt, was a good target for Chang'e-2's flyby mission because
this asteroid is full of fascinating puzzles. The 3-D shape of
Toutatis has been well reconstructed, and its spin state has been
determined, from extensive radar measurements taken at Arecibo and
Goldstone
\cite{Hudson95,Ostro95,Ostro99,Hudson03,Busch10,Busch11,Busch12,Takahashi13}
since 1992 during its closest approaches to Earth, once every four
years. The results indicate that Toutatis is characterized as a
non-principal-axis rotator; this tumbling asteroid rotates about its
long axis with a period of 5.4 days, and the long-axis precession
period is 7.4 days \cite{Busch11} ; this precession may be induced by
the Earth's tides during each close-in flyby \cite{Scheeres07} and
the Yarkovsky-O'Keefe-Radzievskii-Paddack (YORP) effect
\cite{Bottke02} . Moreover, detailed geological features, such as
complex linear structures and concavities, have been revealed in
Hudson et al.'s model \cite{Hudson03} at an average spatial
resolution of $\sim$ 34 m. Hence, the primary issue of concern is
whether the derived radar models accurately represent the real
appearance of Toutatis. Furthermore, some features such as boulders
and regolith on its surface have been speculated to exist in
previous radar models \cite{Hudson03,Busch10} . It was hoped that
this mystery could be solved by obtaining optical images during a
spacecraft's close flyby. Furthermore, small asteroids at the km
scale are believed to have a so-called rubble-pile structure, which
means that they consist of a loose collection of fragments under the
influence of gravity. Therefore, the second major question arises
--- whether Toutatis is also an agglomeration of gravitationally
bound chunks with such a fragile structure.

\begin{figure}
 \vspace{12cm}
\includegraphics{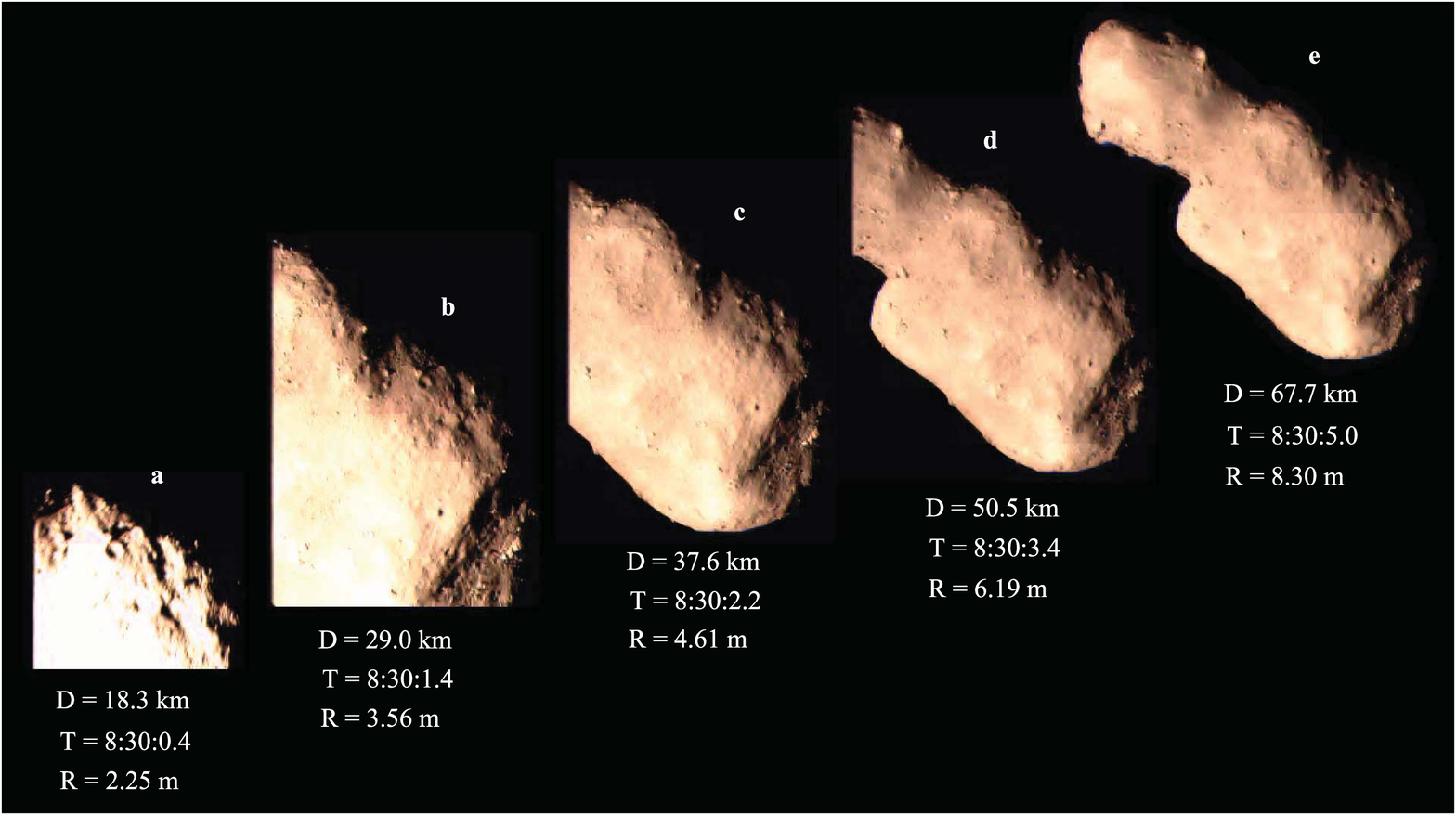}

\caption{\textbf{$\mid$ Outbound images of Toutatis acquired on 13
December, 2012 during Chang'e-2 flyby, indicative of the spacecraft
being away from the asteroid (from a to e).} The left side of
Toutatis is blocked by the solar panel in images a-d.The imaging
distance (D), epoch of flyby (T, UTC) and resolution of each image
(R) are shown for each snapshot, where the distance error for a is
1.1 km. The resolution of each image is linear with the distance.
\label{fig_all}} .

\end{figure}

The optical images of Toutatis were obtained using one of the
onboard engineering cameras, which were originally designed to
monitor the deployment of solar panels on the spacecraft and to
photograph other objects (such as the Earth and Moon) in space. The
camera that is capable of color imaging has a lens with a focal
length of 54 mm and a 1024-by-1024-pixel CMOS detector, and it has a
field of view (FOV) of 7.2$^{\circ}$ by 7.2$^{\circ}$. A solar panel
shields the camera from sunlight to help minimize stray light. The
FOV was partly occulted by the solar panel, and thus some of the
images were cut off from the early gallery. As part of the imaging
strategy, Chang'e-2 maneuvered to adjust the camera's optical axis
to lie antiparallel to the direction of the relative velocity vector
between the probe and Toutatis. Imaging during the inbound
trajectory was not ideal because there was a large
Sun-Toutatis-Chang'e-2 phase angle of $\sim 143.6^{\circ}$.
Therefore, Chang'e-2 completed an attitude adjustment approximately
an hour ahead of the flyby epoch to prepare for the outbound
imaging. The overall imaging time was approximately 25 minutes, and
the interval for each snapshot was 0.2 seconds. During the mission,
the spacecraft collected more than 400 useful images totaling 4.5 GB
of data. Figure \ref{fig_all} shows several outbound images of
Toutatis. Panel e, which was the first panoramic image, was taken at
a distance of $\sim 67.7$ km at a resolution of $\sim 8.30$ m (see
Supplementary Figure 1), whereas panel a, which was the third in a
sequence of images, corresponds to an imaging distance of $\sim
18.3$ km at a local resolution of $\sim 2.25$ m, as shown in Fig.
\ref{fig_all}. The present highest-resolution optical image has a
resolution that is significantly finer than the 3.75 m resolution of
the known radar model \cite{Busch12,Takahashi13} , although the radar
images cover nearly the entire surface of Toutatis, whereas the
optical images cover only one side of Toutatis' surface.

\section*{Results}

We tested several methods to directly establish the 3-D model of
Toutatis based on the optical images collected by Chang'e-2.
However, there is no satisfactory outcome because of the narrow
viewing angle and the local high-resolution images. Therefore, 3-D
models developed based on delay-Doppler radar measurements
\cite{Hudson03,Busch12,Takahashi13} were adopted to discern the
attitude of Toutatis in the approaching epoch.

Based on the attitude of the probe and its camera, the relative
position between the asteroid and the probe, and optical images, in
combination with the models of Hudson et al. and Busch et al.
\cite{Hudson03,Busch12,Takahashi13} , which were utilized to match
the spin state of Toutatis (see Supplementary Figure 2), the
direction of the principal axis was measured to be ($250 \pm
5^\circ, 63 \pm5^\circ$) with respect to the J2000 ecliptic
coordinate system \cite{Scheeres00, Hudson03} , and the fundamental
parameters are summarized in Table 1. Our result is in good
agreement with the previous prediction \cite{Busch12} , which
indicates that Toutatis' spin status was greatly changed because of
tidal interaction with the Earth during the 2004 flyby, which had a
closest distance of $\sim$ 0.01 AU \cite{Scheeres07, Busch12} .
Furthermore, we again measured Toutatis' physical size from the
images, and its maximum length and width were estimated to be (4.75
$\times$ 1.95 km) $\pm$10$\%$, respectively (see Supplementary
Section 3).

\subsection*{New Geological Features from Chang'e-2 Observation}

The shape of Toutatis resembles that of a ginger root (Fig.
\ref{fig_all}), where the bifurcated asteroid mainly consists of a
head (small lobe) and a body (large lobe). The two major parts are
not round in shape, and their surfaces have a number of large
facets. In particular, at the end of the body, there appears to be a
huge square concavity, which could be of large-impact origin.
Regarding the asteroid's appearance, craters are more prominent on
Toutatis than on Itokawa. In contrast, we clearly see fewer boulders
on Toutatis, although many smaller blocks may have been missed
because of the resolution of Chang'e-2 flyby images.

In comparison with the radar models \cite{Ostro95,Ostro99,Hudson03} ,
the proximate observations from Chang'e-2's flyby have revealed
several remarkable discoveries concerning Toutatis, among which the
presence of the giant basin at the big end appears to be one of the
most compelling geological features, and the sharply perpendicular
silhouette in the neck region that connects the head and body is
also quite novel. A large number of boulders and several short
linear structures (Fig. \ref{fig_all}) are also apparent on the
surface, although the lack of sharp contact in the radar model may
be an effect of the SHAPE software \cite{Magri07} .

\subsection*{Concavities}

\begin{figure}
 \vspace{12cm}
\includegraphics{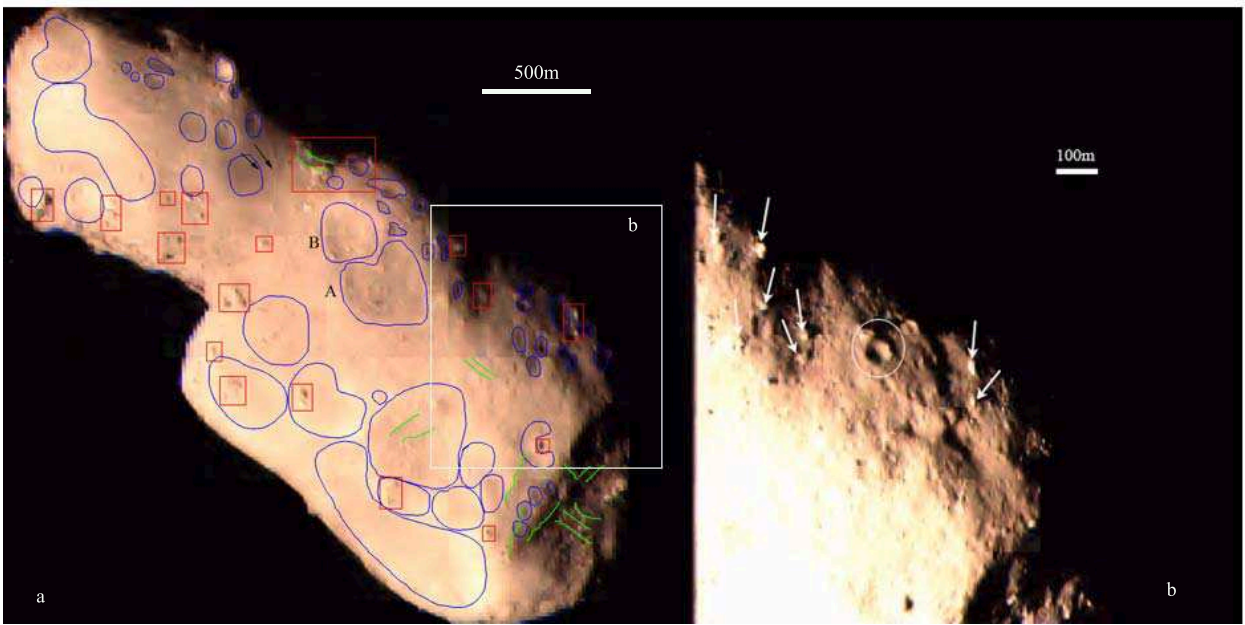}
\caption{\textbf{$\mid$All kinds of geological features on the
Toutatis' surface.} \textbf{(a)} All craters (blue profiles) and
boulders (red squares) are outlined in the panoramic image of
Toutatis. Two craters are closely distributed: the smaller crater
(B) seemed to superimpose on the larger one (A). Green lines
indicate the lineaments. Black arrows point the flow direction of
the fine-grained regolith. \textbf{(b)} The enlarged portion shown
(white box) in the left panel (a). A morphological-integrity crater
can be observed. Tens of boulders are randomly distributed around.
\label{fig_zoom}}
\end{figure}

The giant basin at the big end of Toutatis has a diameter of $\sim$
805 m, and we surmise that one or more impactors may have collided
with the asteroid in this region, thereby gouging out such an
enormous basin. Judging by the contrast of its relatively subdued
relief with large impact basins on other asteroids, Mathilde and
Eros, this large-end depression may be attributable to the internal
structure of the large lobe, with smaller impact features overlying
it. The most significant feature is the ridge around the largest
basin. The wall of this basin exhibits a relatively high density of
lineaments, some of which seem to be concentric to the basin. There
are also several linear structures outside the basin that are
roughly parallel to the basin rim (Fig. \ref{fig_zoom}a). These
ridges are indicative of an internal structure on small bodies
\cite{Prockter02, Thomas12} . Most of the ridges near the largest
basin at the big end are most likely related to the huge stress
energy during impact. The interior material in the deep region would
have been scattered across Toutatis' surface by the large impact
that formed the basin, leading to the fracturing of the asteroid.
Hence, one may speculate that the asteroid most likely has a
rubble-pile internal structure.

More than fifty craters have been identified from the flyby images,
and they have diameters ranging from 36 to 532 m (with an average of
$\sim$ 141 m). $\sim$ 44\% of the craters exceed 100 m in diameter,
which is consistent with Hudson et al.'s model. From the close-up
image, a crater of high morphological integrity (with a diameter of
$\sim$ 62 m) can be clearly observed, whereas tens of boulders (with
lengths of $\sim$ 20-30 m) are randomly distributed nearby (Fig.
\ref{fig_zoom}b). Two concavities are closely carved into the
surface: the larger (with a diameter of $\sim$ 368 m) has a
relatively vague outline, whereas the smaller (with a diameter of
$\sim$ 259 m) has a relatively clear edge. Based on their
appearance, we suspect that the relatively smaller, young crater
(marked with the letter B) is superimposed on the larger, ancient
crater (marked with the letter A) (Fig. \ref{fig_zoom}a).

Although the images show that parts of the observed surface areas
have a high degree of exposure, the topographical features of the
craters were identified based on visual inspection and automatic
annotation. Therefore, we identified craters on the large and small
lobes individually, including vaguely shaped candidate craters on
the highly exposed areas, for the investigation of the cumulative
and relative size-frequency distributions (see Supplementary Section
4) \cite{Zou13} . Figure \ref{fig_fre} indicates that the surfaces of
the two lobes are very likely to share a similar cratering history
because their relative size-frequency profiles for craters
\cite{Arvidson79} look alike throughout a range of diameters.
Furthermore, the craters on the two lobes are not in saturation
equilibrium, as the two profiles do not tend toward stability with
increasing crater diameter (Fig. \ref{fig_fre}a). An abundance of
craters would suggest that Toutatis suffered from many impacts of
interplanetary projectiles in the past.

\subsection*{Boulders}
From the flyby images, more than thirty boulders can be clearly
discerned. They have lengths that range from 12 to 81 m, with an
average of $\sim$ 32 m. Approximately 90\% of the boulders are less
than 50 m in diameter. The largest boulder, along with other large
boulders, is located near the neck region (Fig. \ref{fig_zoom}a). It
is believed that a large ejected fragment is empirically associated
with a large crater \cite{Gault63, Thomas01} . The boulders are
suggested to be those non-escaping ejecta that re-impact onto the
asteroid's surface within a short timescale \cite{Scheeres98} .
Recently, the Goldstone radar images of 12 December 2012 have also
suggested that several 10-m-scale bright features may be boulders
\cite{Busch12} . However, the optical images, although they were
collected on only one side of the asteroid, have an advantage over
the radar models \cite{Hudson03,Busch12} in uncovering local fine
features on the surface of Toutatis.

\subsection*{Regolith}
Two craters in the neck area are indicative of the granular flows in
the inner flanks. The upper regolith near the craters is fine
grained, whereas the lower is relatively coarse, and very many
fragments are present (Fig. \ref{fig_zoom}a). Such a structure may
indicate the redistribution of regolith via downslope or resurfacing
processes, which may be related to the inclined terrain and the
orientation of gravity \cite{Scheeres98} . The results are in good
agreement with those of polarimetric observations \cite{Mukai97,
Hudson98} , thereby indicating that Toutatis is covered with fine
regolith made up of light-transmitting materials. The same features
have also been reported on Lutetia's surface \cite{Vincent12} .
Furthermore, the characteristics once again suggest that Toutatis'
surface may consist of a fine granular or regolith layer with the
porosity of lunar soils \cite{Ostro99} . The latest results imply
that this asteroid may bear an undifferentiated L-chondrite surface
composition \cite{Reddy12} .

There is additional evidence that supports the existence of regolith
on Toutatis' surface. The average thermal inertia for Toutatis may
have a low profile, according to an empirical relation between the
thermal inertia and the effective diameter \cite{Delbo07} . For a
non-principal-axis rotator such as Toutatis, the global
redistribution of regolith that results from lofting may also be one
of the factors that contributes to the low surface thermal inertia
\cite{Delbo07} . A considerable number of mm-sized grains may
constitute Toutatis' regolith (Fig. \ref{fig_zoom}b), and the
regolith depth on the surface might range from several centimeters
to several meters. Clearly, Toutatis' thermal inertia seems to be
much less than that of Itokawa \cite{Fujiwara06} ; thus, the terrain
of Toutatis is a bit smoother than that of Itokawa, and the boulders
on Toutatis' surface are less numerous than those on Itokawa's
surface, which is mostly embedded with numerous pebbles and gravel
in all regions.

\subsection*{Linear Structures}

Several types of linear structures, which are primarily composed of
troughs and ridges that appear to be common for small asteroids
\cite{Thomas10,Robinson02} , can be easily observed on Toutatis'
surface. Notable linear structures are apparent in the radar models
\cite{Hudson03} ; however, the flyby images do not identify any
linear features that have been previously reported because the
relevant part of the surface was invisible to Chang'e-2. The present
linear structures, which are similar to the short linear structures
of Itokawa, are clearly observed \cite{Demura06} to have a length of
120-330 m in the visible regions of the images. The troughs, which
are similar in appearance to groove-like features, are primarily
scattered over the small lobe; they have an average length of $\sim$
170 m, and they are approximately orientated along the long axis of
the asteroid (Fig. \ref{fig_zoom}a). Generally, troughs are related
to extensions resulting from tensile stress in the plane of the
surface \cite{Thomas12} . Consequently, they may be produced by the
tensile stresses that arise from nearby impact cratering or other
geological processes.

\section*{Discussion}
As mentioned above, the existence of the giant basin and its
surroundings provides direct evidence that Toutatis is likely to be
a rubble-pile asteroid. In this case, the asteroid could reassemble
itself into a weak aggregate of large fragments via a heavy impact
or many smaller impacts; in this manner, the large interior voids
could absorb the collision energy and further resist huge collisions
in the formation process. Furthermore, a vast majority of S-type
asteroids appear to possess rubble-pile structures with an average
porosity of $\sim$ 15\% - 25\% \cite{Britt02} . Recent investigations
have demonstrated that the typical bulk density of L ordinary
chondrites is $\sim$ 3.34 g cm$^{-3}~~$ \cite{Britt02,Reddy12} .
Assuming a density of 2.1 - 2.5 g cm$^{-3}~~$ \cite{Scheeres98} ,
Toutatis may have a porosity in the range $\sim$ 25\% to $\sim$
37\%, thereby implying that it may be an intermediate body between
Eros ($\sim$ 20\% \cite{Britt02}) and Itokawa ($\sim$ 41\%
\cite{Fujiwara06}) with respect to surface features. Moreover,
strong evidence from N-body simulations indicates that two km-sized
objects with rubble-pile structures would produce Toutatis-like
objects \cite{Richardson02} . In summary, we may conclude that
Toutatis is not a monolith but most likely a coalescence of
shattered fragments.

The bifurcated configuration means that Toutatis is comprised of two
major lobes, similar to Itokawa, thereby implying that Toutatis is a
contact binary, which is an asteroid type that may constitute more
than 9\% of the NEA population. Km-sized contact binaries, such as
4486 Mithra and 4769 Castalia, share an irregular, significantly
bifurcated shape \cite{Benner05} . The following question then
arises: how do these contact binaries form?

Several formation mechanisms have been proposed to produce such
bifurcated configurations. The first scenario supposes that the two
major lobes were two separate objects, which were moving at a
sufficiently low relative speed to yield a contact binary
\cite{Fujiwara06} . The difference in the orientations of head and
body may be indicative of their diverse origins and further suggests
that the two parts were once detached. The second scenario includes
the YORP and binary YORP (BYORP) effects, which are likely to form
contact binaries \cite{Steinberg11} , especially for those objects
with slow rotational periods. In this scenario, two components,
which may be major fragments of a parent body, suffer re-impact and
recombination, thereby leading to the formation of a bifurcated
configuration. The third scenario suggests that catastrophic
collisions may have gouged out the giant basin and a number of
craters on Toutatis' surface. The boulders near three potential arc
depression terrains (Figure \ref{fig_zoom}a) imply that the head may
have experienced severe impacts. Last but not least, tidal
disruption from a close encounter with a terrestrial planet is also
likely to play a role in the formation of such a contact binary. In
short, the above-mentioned scenarios may have actually occurred
during Toutatis' history. Hence, we infer that Toutatis may be a
reconstituted contact binary.

The Chang'e-2 flyby has revealed several new discoveries regarding
Toutatis, e.g., the giant basin at the big end, the sharply
perpendicular profile about the neck region, and direct images of
distributed boulders and surface regolith. It is the first time that
geological features on Toutatis' surface have been observed so
clearly. All these observations suggest that Toutatis may bear a
rubble-pile structure, similar to other S-type asteroids, such as
Itokawa. Moreover, the close-up imaging results for the asteroid
provide further indication that the bifurcated configuration that is
related to the formation of contact binaries appears to be quite
common among the NEA population. The observations of Chang'e-2 have
significantly improved our understanding of the formation and
evolution of these building blocks of the Solar System.

\section*{Methods}

To adapt to Chang'e-2's characteristics of high control accuracy and
the wide FOV of 7.2$^{\circ}$ by 7.2$^{\circ}$ of the engineering
camera, a close-approach strategy and a specific imaging design were
developed. As shown in Supplementary Figure 1, L, S, and $\Delta t$
represent the minimum distance of rendezvous, the imaging distance,
and the time interval between two sequential images, respectively.
The camera optical axis pointed in the direction of the relative
velocity between the probe and Toutatis, which remained nearly
unchanged during the flyby. After a short time at the closest
approach, the camera captured Toutatis; the asteroid was contained
in the FOV, but the imaging size decreased as the departure distance
increased. By solving the geometric relation between Image 2 and
Image 3, the distances L and S and the resolution of each image were
determined for the Chang'e-2 observations.

Taking into account the various characteristics of the specific
approach design and imaging strategy for the flyby, a physical size
of (4.75 $\times$ 1.95 km) $\pm$10$\%$ for the illuminated region of
the asteroid surface was statistically estimated from the sequence
of images recorded by the CMOS detector during the flyby. The
determined size is in good agreement with that of the radar model
\cite{Hudson03} (see Supplementary Sections 1 and 3).

The attitude of a rigid body in space is determined by its rotations
around three axes of an orthogonal coordinate system. To define the
attitude of Toutatis in each optical image (Supplementary Figure
2a), the three axes $l_1$, $l_2$, and $l_3$ are defined as follows:
$l_1$ and $l_2$ extend through Toutatis' centroid in the images
along the directions of the long axis and short axis, respectively,
and are perpendicular to each other. $l_3$ is perpendicular to the
imaging plane and constitutes a right-handed coordinate system with
$l_1$ and $l_2$.

To match its shape to the optical images, we rotated the model of
Busch et al. in intervals of $5^{\circ}$ for each Euler angle around
the three axes of the body-fixed coordinate system in space. Next,
we chose three criteria of the orientation of the asteroid's
long-axis, the ratio of length to width, and the obvious
topographical feature at the joint of the two lobes of Toutatis ---
to ensure that the radar model agreed with the optical images from
Chang'e-2, as shown in Supplementary Figure 2a. The most approximate
attitude was finally obtained by rotating the radar model and
comparing it with the optical image, as shown in Supplementary
Figure 2b. The coordinate transformation from the asteroid's
body-fixed system to the J2000.0 equatorial coordinate system were
expressed in terms of three components of Euler angles. The
directions of the principal axes were then obtained according to the
attitude of the radar model.

The craters and other geological features were identified via both
visual inspection and automatic annotation. In this case, dozens of
low-lying regions were noted, and their positions, contours, and
sizes were obtained using surface-structure-analysis techniques. The
feature points on the edge of each crater were labeled manually
based on the treatment of the imaging process and the radar model.
After the points were marked, a piecewise-quadratic curve was fitted
to them, and closed curves were obtained as contour profile. The
sizes of the craters (and boulders) were calculated based on the
resolution for each image and the fitted contour profiles.

{}

\section*{}
\textbf{Supplementary Information} is available in the online
version of the paper.

\section*{}
\textbf{Acknowledgements} The authors thank Michael W. Busch for his
constructive comments that greatly helped to improve the original
content of this manuscript. We thank the whole Chang'e-2 team to
make the mission a success. We appreciate B. Yang for target
observations before the flyby mission. We are grateful to R.P.
Butler for English improvement of this manuscript. This research was
supported by National Science and Technology Major Project, National
Natural Science Foundation of China, the Innovative and
Interdisciplinary program by CAS, and Key Laboratory of Planetary
Sciences (2013DP173302), CAS. JYL's contribution to this work is not
supported by any funds.

\section*{}

\textbf{Authors Contributions}~J. C. H. was the leader of the
asteroid mission. J. H. J. wrote the manuscript with contributions
from other co-authors. J.C.H., P.J.Y. and T.X.Z. were responsible
for the overall project design of the flyby mission to Toutatis,
imaging strategy, and proposed the method of fundamental parameter
analysis. H.B.Z., J.H.J., S.C.H., S.W., Y.H.Z., Y.J. and L.L.Y.
contributed to ground-based observation, refined the
trajectory solution for Toutatis, completed data analysis, prepared
for figures and wrote the manuscript. L.Z.M, H.H. and W.R. took charge of
overall design of probe system and data analysis. X.L.W. was
responsible for controlling the attitude and the spacecraft's orbit,
and contributed to the engineering parameters analysis. J.Y.,
C.L.L., Y.L.Z. and X.J.J. took charge of data collection, data
pretreatment and the ground-based observation. Y.L. and M.H.Z.
partly contributed to contents and prepared for figure. D.Q.
contributed to the target selection and the design of transfer
orbit. W.Z., P.L. and X.L.T. were in charge of the algorithm of optical
images and obtained the image information. S.L. was responsible for
imaging process. C.N.H. developed the optical camera and determined
the parameters of the camera. W.H.I and J.Y.L. contributed ideas and
discussions in preparation for the manuscript. J.M.B. partly
participated in the target observation. All authors contributed to
the manuscript.

\section*{}
\textbf{Authors information} The authors declare no competing financial interests.
To whom correspondence should be addressed. E-mail:jijh@pmo.ac.cn.

\bibliography{scibib}

\begin{thebibliography}{}

\bibitem[1]{Huang13}    Huang, J. C. et al. The engineering parameters analysis of (4179) Toutatis flyby mission of Chang'e-2. {\it Science China Technological Sciences} {\bf 43}, 596-601 (2013).
\bibitem[2]{Hudson95}   Hudson, R. S. \& Ostro, S. J. Shape and non-principal axis spin state of Asteroid 4179 Toutatis.{\it Science} {\bf 270}, 84-86 (1995).
\bibitem[3]{Ostro95}    Ostro, S. J. et al. Radar images of Asteroid 4179 Toutatis. {\it Science} {\bf 270}, 80-83 (1995).
\bibitem[4]{Ostro99}    Ostro, S. J. et al. Asteroid 4179 Toutatis: 1996 radar observations. {\it Icarus} {\bf 137}, 122-139 (1999).
\bibitem[5]{Hudson03}   Hudson, R. S., Ostro, S. J. \& Scheeres, D. J. High-resolution model of Asteroid 4179 Toutatis. {\it Icarus} {\bf 161}, 346-355 (2003).
\bibitem[6]{Busch10}    Busch, M. W. et al. Determining asteroid spin states using radar speckles. {\it Icarus} {\bf 209}, 535-541 (2010).

%%\bibitem[7]{Busch11}    Busch, M. W. et al. Twenty years of Toutatis, EPSC-DPS Joint Meeting, Nantes, France, (2011).
%%\bibitem[8]{Busch12}    Busch, M. W. et al. Internal structure of 4179 Toutatis, 2012 AGU Fall Meeting, San Francisco, CA, USA. %%(2013).

%%Vol. 6, EPSC-DPS2011-297,

\bibitem[7]{Busch11}    Busch, M. W. et al. Twenty years of Toutatis. {\it EPSC-DPS2011} {\bf 6}, 297 (2011).

\bibitem[8]{Busch12}    Busch, M. W. et al. Internal structure of 4179 Toutatis. 2012 AGU Fall Meeting, P31A-1873 (2012).

\bibitem[9]{Takahashi13}Takahashi, Y., Busch, M. W. \& Scheeres, D. J. Spin State and Moment of Inertia Characterization of 4179 Toutatis, {\it AJ} {\bf 146}, 95 (2013).

\bibitem[10]{Scheeres07}Scheeres, D. J. Rotational fission of contact binary asteroids. {\it Icarus} {\bf 189}, 370-385 (2007).

\bibitem[11]{Bottke02}  Bottke, W. F., Vokrouhlicky, D., Rubincam, D. P. \& Broz, M. in {\it Asteroids III} (eds. Bottke, W. F. et al.)(Univ. of Arizona Press, Tucson, 2002).
\bibitem[12]{Scheeres00}Scheeres D. J., Ostro S. J., Werner R A, et al. Effects of gravitational interactions on asteroid spin states. {\it Icarus} {\bf 147}, 106-118 (2000).

\bibitem[13]{Magri07} Magri, C. et al. Radar observations and a physical model of Asteroid 1580 Betulia {\it Icaurs} {\bf 186}, 152-177 (2007)
\bibitem[14]{Prockter02} Prockter, L. et al. Surface expressions of structural features on Eros. {\it Icarus} {\bf 155}, 75-93 (2002).
\bibitem[15]{Thomas12}   Thomas, N. et al. The geomorphology of (21) Lutetia: Results from the OSIRIS imaging system onboard ESA's Rosetta spacecraft. {\it Planet. Space Sci.}  {\bf 66}, 96-124 (2012).

\bibitem[16]{Zou13} Zou, X.D. et al. Preliminary analysis of the 4179 Toutatis snapshots of the Chang'e-2 fly-by. {\it Icarus}, in press (2013).
\bibitem[17]{Arvidson79} Arvidson, R. E. et al. Standard techniques for presentation and analysis of crater size-frequency data. {\it Icarus} {\bf 37}, 467-474 (1979).
\bibitem[18]{Gault63}    Gault, D. E., Shoemaker, E. M. \& Moore, H. J. Spray ejected from the lunar surface by meteoroid impact. NASA Technical Document TND-1767, 1-39 (1963).
\bibitem[19]{Thomas01}   Thomas, P. C., Veverka, J., Robinson, M. S. \& Murchie, S. Shoemaker crater as the source of most ejecta blocks on the asteroid 433 Eros. {\it Nature} {\bf 413}, 394-396 (2001).
\bibitem[20]{Scheeres98}Scheeres, D. J., Ostro, S. J., Hudson, R. S., DeJong, E. M. \& Suzuki, S. Dynamics of orbits close to Asteroid 4179 Toutatis. {\it Icarus} {\bf 132}, 53-79 (1998).


\bibitem[21]{Mukai97}    Mukai, T. et al. Polarimetric observations of 4179 Toutatis in 1992/1993. {\it Icarus} {\bf 127}, 452-460 (1997).
\bibitem[22]{Hudson98}   Hudson, R. S. \& Ostro, S. J. Photometric properties of Asteroid 4179 Toutatis from lightcurves and a radar-derived physical model. {\it Icarus} {\bf 135}, 451-457 (1998).
\bibitem[23]{Vincent12}  Vincent, J. B., Besse, S., Marchi, S., Sierks, H. \& Massironi, M. Physical properties of craters on asteroid (21) Lutetia. {\it Planet. Space Sci.} {\bf 66}, 79-86 (2012).
\bibitem[24]{Reddy12}    Reddy, V. et al. Composition of near-Earth Asteroid (4179) Toutatis. {\it Icarus} {\bf 221}, 1177-1179 (2012).
\bibitem[25]{Delbo07}    Delbo', M., dell'Oro, A., Harris, A. W., Mottola, S. \& Mueller, M. Thermal inertia of near-Earth asteroids and implications for the magnitude of the Yarkovsky effect. {\it Icarus} {\bf 190}, 236-249 (2007).

\bibitem[26]{Fujiwara06}  Fujiwara, A. et al. The rubble-pile asteroid Itokawa as observed by Hayabusa. {\it Science} {\bf 312}, 1330-1334 (2006).
\bibitem[27]{Thomas10}   Thomas, P. C. \& Prockter, L. M. in {\it Planetary Tectonics}(eds. Watters, T. R \& Schultz, R. A.) (Cambridge University Press, New York, 2010).
\bibitem[28]{Robinson02} Robinson, M. S., Thomas, P. C., Veverka, J., Murchie, S. L. \& Wilcox, B. B. The geology of 433 Eros. {\it Meteorit. Planet. Sci.} {\bf 37}, 1651-1684 (2002).
\bibitem[29]{Demura06}   Demura, H. et al. Pole and global shape of 25143 Itokawa. {\it Science} {\bf 312}, 1347-1349 (2006).

\bibitem[30]{Britt02}    Britt, D. T., Yeomans, D., Housen, K. \& Consolmagno, G. in {\it Asteroids  III}(eds. Bottke, W. F., Cellino, A., Paolicchi, P., \& Binzel, R. P.)(Univ. of Arizona Press, Tucson, 2002).
\bibitem[31]{Richardson02} Richardson, D. C., Leinhardt, Z. M., Melosh, H. J., Bottke, W. F. Jr. \& Asphaug, E. in {\it Asteroids  III}(eds. Bottke, W. F., Cellino, A., Paolicchi, P., \& Binzel, R. P.)(Univ. of Arizona Press, Tucson, 2002).

\bibitem[32]{Benner05}    Benner, L. A. M. et al. Near-Earth Asteroid 2005 CR37: radar images and photometry of a candidate contact binary. {\it Icarus} {\bf 182}, 474-481 (2005).
\bibitem[33]{Steinberg11} Steinberg, E. \& Sari, R. Binary YORP effect and evolution of binary asteroids. {\it AJ} {\bf 141}, 55 (2011).
%%\bibitem[34]{Harris02}   Harris, A. W. On the slow rotation of asteroids. {\it Icarus} {\bf 156}, 184-190 (2002).
\bibitem[34]{Krivova94}  Krivova, N. V., Yagudina, E. I, \& Shor, V. A. The orbit determination of (4179) Toutatis from optical and radar data. {\it Planet. Space Sci.} {\bf 42}, 741-745 (1994).
\bibitem[35]{Milani10}   Milani, A., \& Gronchi, G. Theory of orbit determination. (Cambridge University Press, Cambridge, UK, 2010).



\end{thebibliography}

\bibliographystyle{Science}

\clearpage

\clearpage

%%Figure 2 caption should be modified
\clearpage

%%Figure 3 caption should be modified
\begin{figure}
 \vspace{12cm}
\includegraphics{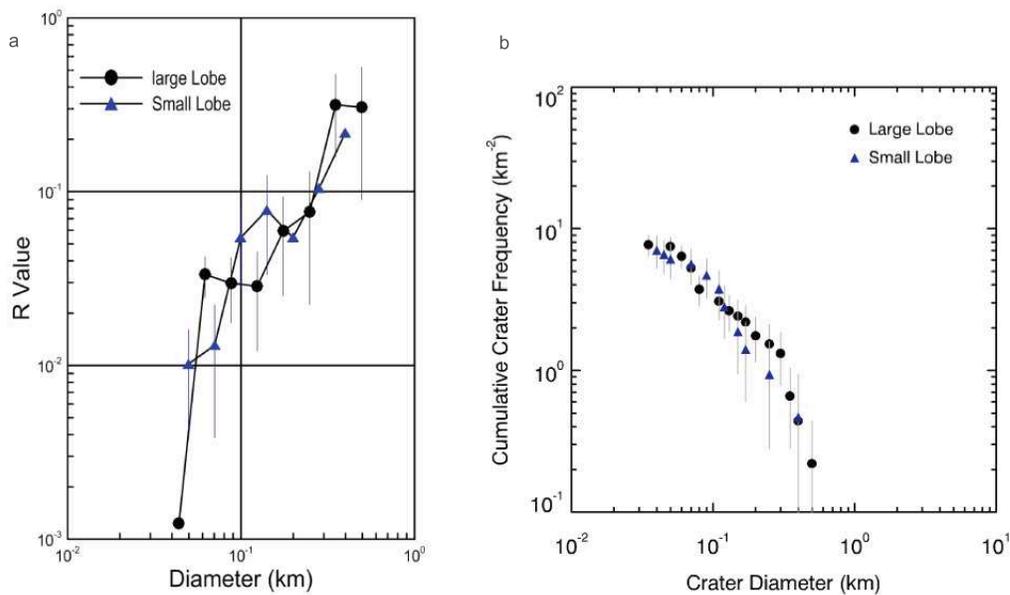}
%%\special{psfile=Fig3.eps  vscale=48 hscale=48 angle=0}
\caption{\textbf{$\mid$(a)The relative size-frequency distribution
for craters on Toutatis. (b) The cumulative  size-frequency
distribution for craters on Toutatis.} In panel (a) R plot was
devised by the Crater Analysis Techniques Working Group
\cite{Arvidson79} to better show the size distribution of craters
and crater number densities for determining relative ages. The
vertical position of the curve is a measure of crater density or
relative age on the Toutatis: the higher the vertical position, the
higher the crater density and the older the surface. In panel (b)
$x$-axis stands for the crater diameter and $y$-axis represents the
crater number larger than corresponding diameter on the investigated
area. CSFD would easily inform that the size of craters distribute
as diameters increase. \label{fig_fre}}
\end{figure}

\clearpage
%\begin{center}
{\bf Table 1. Parameters for Toutatis.}

\begin{tabular}{ll}
\hline Property & Value \\ \hline
Osculating Orbital elements &  $a$=2.5336 AU,  ~$e$=0.6301 \\
  & $i$=0.4466$^\circ$ ,    ~~~~~~$\Omega$=124.3991$^\circ$ \\
  & $\omega$=278.6910$^\circ$ , ~$M$=6.7634$^\circ$  \\
Special type &S (IV) \cite{Hudson03} \\
Size (diameter) Major axes
%% & (x=4.12$\pm$0.24 km, y=1.74$\pm$ 0.10 km, z=1.44$\pm$ 0.09 km)\\
  &(x=4.60$\pm$0.10 km, y=2.29$\pm$0.10 km, z=1.92$\pm$ 0.10 km) \cite{Hudson03}\\
  &(x=4.46$\pm$0.10 km, y=2.27$\pm$0.10 km, z=1.88$\pm$ 0.10 km) \cite{Takahashi13}\\

  Rotational properties & \\
  Rotation   & 5.4 day   \cite{Hudson95, Hudson03}\\
  Precession & 7.4 day   \cite{Hudson95, Hudson03}\\
  Pole position
%%    in the space &$\beta=249.8 \pm 5^\circ$,   $\lambda=62.6\pm5^\circ$\\
      in the space &$\beta=250 \pm 5^\circ$, ~~ $\lambda=63 \pm5^\circ$\\
                &$\beta=54.6^\circ$, ~~~~~~~  $\lambda=60.6^\circ$ \cite{Busch12,Takahashi13}\\
                &$\alpha=258\pm5^\circ$, $\delta=40\pm5^\circ$\\

    Density & 2.5 g cm$^{-3}$    ~~    \cite{Scheeres98}\\
\hline \label{param}
\end{tabular}

Using the data released by the Minor Planet Center\cite{Krivova94,
Milani10} and optical data from the ground-based observational
campaign sponsored by the Chinese Academy of Sciences, we determined
Toutatis' orbit with uncertainties on the order of several
kilometers. Osculating orbital elements were calculated for the
flyby epoch of 13 December 2012 at 8:30 UTC, where $a,~ e, ~i,~
\Omega, ~\omega$, and $M$ are the semi-major axis, the eccentricity,
the inclination, the longitude of the ascending node, the argument
of perihelion, and the mean anomaly, respectively. $\beta, ~\lambda$
and $\alpha,~ \delta$ are the longitudes and latitudes of the long
axis of the asteroid in the J2000.0 ecliptic and equatorial
coordinate systems, respectively.
%\end{center}

\end{document}